\documentclass[11pt,a4paper]{article}

\usepackage{amsmath,amssymb,tikz,hyperref,empheq}
\usepackage{graphicx}
 \usepackage{verbatim}
 \allowdisplaybreaks
\def\be{\begin{equation}}
\def\ee{\end{equation}}
\def\ba#1\ea{\begin{align}#1\end{align}}
\def\bg#1\eg{\begin{gather}#1\end{gather}}
\def\bm#1\em{\begin{multline}#1\end{multline}}
\def\bmd#1\emd{\begin{multlined}#1\end{multlined}}

\def\({\left(}
\def\){\right)}
\def\[{\left[}
\def\]{\right]}

\def \be {\begin{equation}}
\def \ee {\end{equation}}
\def \ba {\begin{array}}
\def \ea {\end{array}}
\def \bea{\begin{eqnarray}}
\def \eea{\end{eqnarray}}

\def\bea{\begin{eqnarray}}
\def\eea{\end{eqnarray}}

\newcommand{\bit}{\begin{itemize}}  \newcommand{\eit}{\end{itemize}}
\newcommand{\ben}{\begin{enumerate}}  \newcommand{\een}{\end{enumerate}}

\long\def\symbolfootnote[#1]#2{\begingroup%
\def\thefootnote{\fnsymbol{footnote}}\footnote[#1]{#2}\endgroup}

\newcommand{\sysu}{{\it School of Physics and Astronomy, Sun Yat-Sen University, 2 Daxue Road, Zhuhai 519082, China}}

\begin{document}
\thispagestyle{empty}
\begin{center}

~\vspace{20pt}

{\Large\bf Holographic Bound of Casimir Effect in General Dimensions}

\vspace{25pt}

Rong-Xin Miao ${}$\symbolfootnote[1]{Email:~\sf miaorx@mail.sysu.edu.cn}

\vspace{10pt}${}$\sysu

\vspace{2cm}

\begin{abstract}
Recently, it has been proposed that holography imposes a universal lower bound on the Casimir effect for 3d BCFTs. This paper generalizes the discussions to higher dimensions. We find Einstein gravity, DGP gravity, and Gauss-Bonnet gravity sets a universal lower bound of the strip Casimir effect in general dimensions. We verify the holographic bound by free theories and $O(N)$ models in the $\epsilon$ expansions. We also derive the holographic bound of the Casimir effect for a wedge and confirm free theories obey it. It implies holography sets a lower bound of the Casimir effect for general boundary shapes, not limited to the strip. Finally, we briefly comment on the impact of mass and various generalizations and applications of our results. 
\end{abstract}

\end{center}

\newpage
\setcounter{footnote}{0}
\setcounter{page}{1}

\tableofcontents

\section{Introduction}

With the maturity of experimental techniques and potential applications in nanotechnology \cite{Mohideen:1998iz, Bressi:2002fr, Klimchitskaya:2009cw}, the Casimir effect \cite{Casimir:1948dh} has attracted increasing attention  \cite{Plunien:1986ca, Bordag:2001qi, Milton:2004ya}. The Casimir effect also assumes a crucial theoretical significance in cosmology \cite{Wang:2016och}, QCD \cite{Vepstas:1984sw}, and worm-hole physics \cite{Morris:1988tu, Maldacena:2020sxe}. By definition, Casimir energy is the lowest energy of a quantum system with a boundary. Since energy should be bounded from the below, the Casimir effect is expected to have a fundamental lower bound. Inspired by the Kovtun-Son-Starinet (KSS) bound \cite{Kovtun:2004de}, \cite{Miao:2024gcq} proposes that holography imposes a lower bound of the Casimir effect. The holographic bound passes the tests of free BCFTs, Ising model and $O(N)$ models with $N=2,3$ in three dimensions. Notably, unlike the KSS bound depending on the specific holographic models \cite{Brigante:2007nu, Kats:2007mq, Brigante:2008gz}, holography establishes a universal lower bound for Casimir effect \cite{Miao:2024gcq}. The initial work \cite{Miao:2024gcq} mainly focuses on 3d BCFTs. We generalize the discussions to higher dimensions in this paper. 

Let us consider the Casimir effect of a strip in a d-dimensional flat space
\begin{eqnarray}\label{Tij strip} 
\langle T^i_{\ j} \rangle_{\text{strip}}= \frac{\kappa_1}{L^d}\text{diag}\Big(1, -(d-1),1,..,1\Big),
\end{eqnarray}
where $\kappa_1$ is the dimensionless Casimir amplitude, and $L$ is the strip width. Note that we impose the same boundary conditions on the two plates. Following \cite{Miao:2024gcq}, we propose the Casimir amplitude ratio to the norm of displacement operator has a lower bound in general dimensions
\begin{eqnarray}\label{Casimir bound}
(\frac{-\kappa_1}{C_D})\ge\frac{-2^{d-2} d \pi ^{d-\frac{1}{2}}\Gamma \left(\frac{d-1}{2}\right) \Gamma \left(\frac{1}{d}\right)^d}{\Gamma (d+2) \left(d \Gamma \left(\frac{1}{2}+\frac{1}{d}\right)\right)^{d} }.
\end{eqnarray}
The displacement operator $D(y)$ describes the violation of space translation invariance normal to the boundary \cite{Billo:2016cpy}
\begin{eqnarray}\label{DT}
\nabla_i T^{ij}=\delta(x)D(y) n^j,
\end{eqnarray}
and has a positive norm $C_D\ge 0$ defined by the two-point function \cite{Billo:2016cpy}
\begin{eqnarray}\label{Zamolodchikov norm} 
\langle D(y) D(0) \rangle =\frac{C_D}{|y|^{2d}}.
\end{eqnarray}
A quick way to derive the holographic bound (\ref{Casimir bound}) is by using Einstein gravity with the minimal tension $T\to -(d-1)$, where the corresponding $\kappa_1$ and $C_D$ can be found in \cite{Miao:2024ddp}. In this paper, we investigate the holographic lower bound (\ref{Casimir bound}) and verify its universality by studying Dvali-Gabadadze-Porrati (DGP) gravity \cite{Dvali:2000hr}, Gauss-Bonnet (GB) gravity and GB-DGP gravity. We test it with free theories and the $O(N)$ model in the $\epsilon$ expansions. See Table. \ref{table1 Casimir bounds} for a glance for 3d and 4d BCFTs. We remark that the ratios $(-\kappa_1/C_D)$ coincide for $d=2$ and approach zero for $d\to \infty$ for all BCFTs. See Fig. \ref{ratios HD} for example. 
\begin{table}[ht]
\caption{$(-\kappa_1/C_D)$ for various 3d and 4d BCFTs}
\begin{center}
    \begin{tabular}{| c | c | c | c |  c | c | c | c| c| c|c| }
    \hline
     &Scalar&Fermion& Maxwell & Holography\\ \hline
3d & $\frac{-\pi  \zeta (3)}{2}\approx -1.89$& $\frac{-3\pi  \zeta (3)}{8}\approx -1.42 $ & $\times$  & $\frac{-\pi ^{5/2} \Gamma \left(\frac{1}{3}\right)^3}{108 \Gamma \left(\frac{5}{6}\right)^3}\approx -2.17$\\ \hline
4d&$\frac{-\pi ^6}{720}\approx -1.34$& $\frac{-7 \pi ^6}{8640}\approx -0.78$ & $\frac{-\pi ^6}{4320}\approx -0.22$ & $-\frac{\pi ^4 \Gamma \left(\frac{5}{4}\right)^4}{15 \Gamma \left(\frac{3}{4}\right)^4}\approx -1.94 $\\ \hline
    \end{tabular}
\end{center}
\label{table1 Casimir bounds}
\end{table}

We also investigate the holographic bound of the Casimir effect for a wedge, which is the simplest generalization of a strip. The wedge Casimir effect takes the following form in a ground state
\begin{eqnarray}\label{Tij wedge} 
\langle T^i_{\ j} \rangle_{\text{wedge}}=\frac{f(\Omega)}{r^d}\text{diag}\Big(1, -(d-1),1,..,1\Big),
\end{eqnarray}
where $f(\Omega)$ is the Casimir amplitude, $\Omega$ is the opening angle  of wedge, and $r$ is the distance to the corner of wedge. Similar to the strip, we propose the ratio of Casimir amplitude to the norm of displacement operator has a lower bound set by holography 
\begin{eqnarray}\label{wedge bound} 
(\frac{-f(\Omega)}{C_D})\ge \lim_{T\to -(d-1)} (\frac{-f(\Omega)}{C_D})_{\text{holo}}, \ \text{for} \ 0<\Omega\le \pi.
\end{eqnarray}
In the following, we derive the holographic lower bound by Einstein's gravity and verify it by free BCFTs. See Fig. \ref{wedge bound 3d} and Fig. \ref{wedge bound 4d} for the ratios $(-f(\Omega)/C_D)$ for various 3d and 4d BCFTs. The holographic lower bound (\ref{wedge bound}) is obtained analytically for $d=2, 4$ and numerically for general dimensions. It suggests holography sets a lower bound of the Casimir effect for general boundary shapes, not limited to the strip. 

In summary, in comparison to \cite{Miao:2024gcq}, this paper offers new discussions on higher dimensions, more general gravity duals (GB and GB-DGP gravity), and provides exact lower bounds for the wedge Casimir effect, as detailed in eq.(\ref{sect6: ratio 2d 4d}), as well as in Figures \ref{wedge bound 3d} and \ref{wedge bound 4d}. 

The paper is organized as follows. In section 2, we study the holographic bound of the strip Casimir effect for DGP gravity in general dimensions. We show detailed calculations. We find two phases depending on the DGP couplings. The normal phase can continuously transform into that of Einstein's gravity and imposes a universal lower bound of the Casimir effect. The singular phase cannot continuously transform into that of Einstein's gravity and is always larger than the holographic lower bound. We generalize the discussions to GB gravity and GB-DGP gravity in section 3 and section 4, respectively. Section 5 tests the holographic lower bound of the strip Casimir effect. Section 6 discusses the holographic lower bound of the Casimir effect for a wedge. Finally, we conclude with some open problems in section 7. Appendix A and Appendix B derive the ghost-free condition and the norm of displacement operator for GB-DGP gravity.

\section{Holography I: DGP gravity}

In this section, we study the holographic bound of the Casimir effect of a strip for DGP gravity in general dimensions. Depending on the DGP parameters, there are two phases. The normal phase yields a universal lower bound of the Casimir effect. On the other hand, the singular phase obeys but cannot saturate the lower bound of the Casimir effect. 

We start with the DGP gravity 
\begin{eqnarray}\label{sect2: DGPgravity}
I_{\text{DGP}}=\int_N d^{d+1}x \sqrt{|g|} \Big(R +d(d-1)\Big)+2\int_Q d^dy \sqrt{|h|} \Big(K-T +\lambda \mathcal{R} \Big),
\end{eqnarray}
where $R$ is Ricci scalar in bulk $N$, $K$ and $\mathcal{R}$ are extrinsic curvature and Ricci scalar on the brane $Q$, $T$ denotes the brane tension, and $\lambda$ is the DGP coupling. For simplicity, we have set the Newton constant to be $16\pi G_N=1$ and the AdS radius to be $l=1$. We choose Neumann boundary condition (NBC) \cite{Takayanagi:2011zk} on the brane $Q$
\begin{eqnarray}\label{sect2: NBC}
K^{ij}-(K-T+\lambda \mathcal{R}) h^{ij}+2 \lambda \mathcal{R}^{ij}=0,
\end{eqnarray}
where the brane tension is parameterized as \cite{Miao:2023mui}
 \begin{eqnarray}\label{sect2: Tension}
T=(d-1) \tanh(\rho)-(d-1)(d-2)\lambda \text{sech}^2(\rho). 
 \end{eqnarray} 
 Let us explain the origin of this parameterization. The local region near any smooth boundary can be approximately a half-space. Thus, the gravity dual of the near-boundary region is given by that of half space
\begin{eqnarray}\label{sect2: AdS}
&&\text{metric}:\ \ ds^2=\frac{dz^2-dt^2+dx^2+\sum_{a=1}^{d-2}dy_a^2+ O(z^2)}{z^2}, \\
&&\text{brane for half space } x\ge 0:\ \  x=-\sinh(\rho) z+O(z^2), \label{sect2: QL}\\
&&\text{brane for half space } x\le L:\ \ x-L=\sinh(\rho) z+O(z^2). \label{sect2: QR}
\end{eqnarray}
Then the NBC (\ref{sect2: NBC}) at leading order yields (\ref{sect2: Tension}). We focus on positive DGP gravity to be ghost-free  \cite{Miao:2023mui}
 \begin{eqnarray}\label{sect2: positive DGP}
\lambda\ge 0.
 \end{eqnarray} 
The holographic norm of displacement operator is given by \cite{Miao:2023mui,Miao:2018dvm} for $\rho\le 0$
\begin{eqnarray}\label{sect2: CD d}
C_D=\frac{2 (d-1) \pi ^{\frac{1}{2}-\frac{d}{2}} \Gamma (d+2) (-\text{csch}(\rho ))^{-d}}{\Gamma \left(\frac{d+1}{2}\right) \left(\frac{d \sinh (\rho ) \cosh (\rho ) (-\coth (\rho ))^{3-d}}{2 (d-2) \lambda +\coth (\rho )}-\, _2F_1\left(\frac{d-1}{2},\frac{d}{2};\frac{d+2}{2};-\text{csch}^2(\rho )\right)\right)}.
\end{eqnarray}
The one with $\rho\ge 0$ can be obtained by analytical continuation. For example, we have 
\begin{equation}\label{sect2: CD 3d 4d}
C_D=\begin{cases}
 \frac{32}{\pi  \left(\frac{2 \lambda }{2 \lambda  \sinh (\rho )+\cosh (\rho )}+2 \tan ^{-1}\left(\tanh \left(\frac{\rho }{2}\right)\right)+\frac{\pi }{2}\right)},&\
\text{for} \ d=3,\\
  \frac{120 e^{-\rho } (4 \lambda  \sinh (\rho )+\cosh (\rho ))}{\pi ^2 (2 \lambda  \tanh (\rho )+2 \lambda +1)},&\
\text{for} \ d=4.
\end{cases}
\end{equation}
$C_D\ge 0$ together with $\lambda\ge 0$ yields the constraints
\begin{eqnarray}\label{sect2: constraints1}
&& 0\le \lambda , \ \ \ \ \ \ \ \ \ \ \ \ \ \ \ \  \ \ \text{for } \ \rho \ge 0,\\
&& 0\le  \lambda \le -\frac{\coth (\rho )}{2(d-2)}, \ \text{for } \ \rho \le 0. \label{sect2: constraints2}
\end{eqnarray}
For $0\le \lambda\le 1/2(d-2)$, the above constraints are automatically satisfied for arbitrary $\rho$. That is because $-\frac{\coth (\rho )}{2(d-2)}\ge \frac{1}{2(d-2)}$ for $\rho\le 0$. While for $\lambda> 1/2(d-2)$, (\ref{sect2: constraints2}) sets a lower bound $\rho> -\coth ^{-1}(2(d-2) \lambda )$. Thus, $\lambda=1/2(d-2)$ is a critical DGP parameter. Below, we will show $\lambda=1/2(d-2)$ is the phase-transition point for normal and singular phases. 

\subsection{General discussions}

Now, let us study the gravity dual of a strip with width $L$. The vacuum of a trip is dual to  the AdS soliton \cite{Fujita:2011fp}
\begin{eqnarray}\label{sect2: AdS soliton}
ds^2=\frac{\frac{dz^2}{h(z)}+h(z)d\theta^2-dt^2 +\sum_{a=1}^{d-2}dy_a^2}{z^2},
\end{eqnarray}
where $h(z)=1-z^d/z_h^d$. To avoid conical singularities, we choose the angle period in bulk as
\begin{eqnarray}\label{sect2: angle period}
\beta=\frac{4\pi}{ |h'(z_h)|}=\frac{4\pi z_h}{d}.
\end{eqnarray}
The strip is given by $0\le \theta\le L$ on the AdS boundary $z=0$. By applying the holographic renormalization \cite{deHaro:2000vlm}, we derive the holographic energy density $T_{tt}=-1/z_h^d$. Compared with the Casimir effect (\ref{Tij strip}), we get the holographic Casimir amplitude
\begin{eqnarray}\label{sect2: kappa1}
\kappa_1=\frac{L^d}{z_h^d},
\end{eqnarray}
where the strip width $L$ will be determined soon. Without loss of generality, we set $z_h=1$ below. We label the embedding function of brane $Q$ as
\begin{eqnarray}\label{sect2: soliton Q}
\theta=S(z).
\end{eqnarray}

\begin{figure}[t]
\centering
\includegraphics[width=9cm]{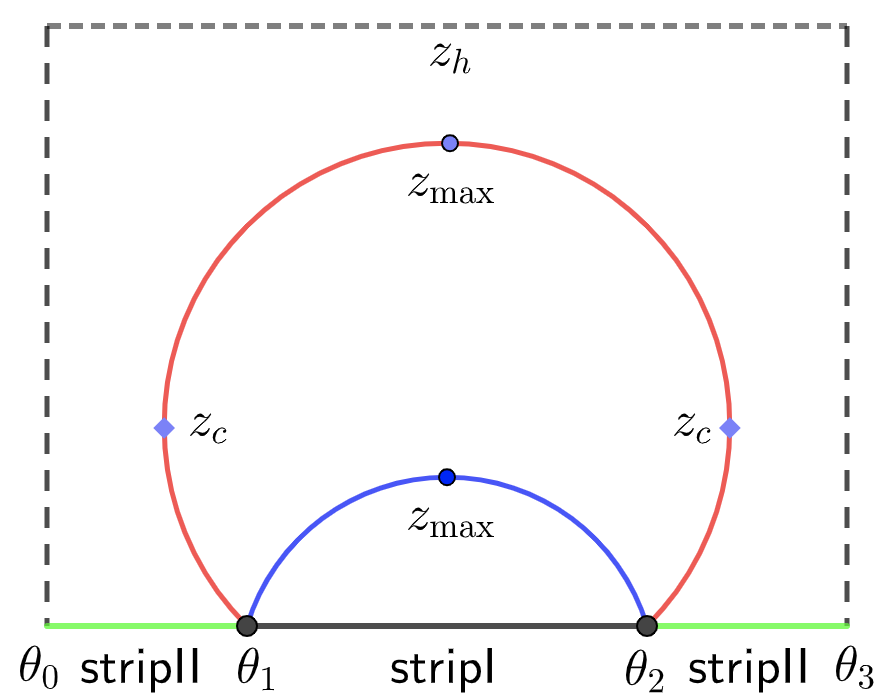}
\caption{Geometry of holographic strip: a portion of AdS soliton. The region between red/blue curves (branes) and the black line is the bulk dual of strip I with negative brane tension $T\le 0$; its complement in bulk is the gravity dual for strip II (green line) with $T> 0$. The gravity dual of strip II contains the `horizon' $z=z_h$. The angle $\theta$ should be periodic to remove the conical singularity on it. Thus, the green lines for strip II are connected. Without loss of generality, we focus on strip I with $T\le 0$. We have $(\rho<0)$ and $(\rho>0)$ for the blue and red curves, respectively. $z_{\text{max}}$ and $z_c$ denote the turning points with $\theta'(z_{\text{max}})=\infty$ and $\theta'(z_{c})=0$.}
\label{holo strip}
\end{figure}

Let us first study the case with negative brane tension $T\le  0$, whose complement can give the other case with $T>0$. It is the typical trick used in \cite{Fujita:2011fp}. Let us explain more. See Fig. \ref{holo strip} for the geometry of the holographic dual of the strip. The region between red/blue curves (branes) and the black line is the bulk dual of strip I with $T\le 0$; its complement in bulk is the gravity dual for strip II (green line) with $T> 0$. As shown in Fig. \ref{holo strip}, the gravity duals of strip I and strip II share the same EOW branes (blue curve or red curve, depending on the theory parameters). As a result, the NBCs (\ref{sect2: NBC}) for strips I and II take the same values. However, the extrinsic curvatures $K_{ij}$ flip signs, while the induced metric $h_{ij}$ and intrinsic Ricci tensor $\mathcal{R}_{ij}$ remain invariant when crossing the branes. In the viewpoint of NBC (\ref{sect2: NBC}), $T$ and $\lambda$ change signs equivalently when crossing the branes. Note that the bulk dual for strip II contains the `horizon' $z=z_h$. To remove the conical singularity at $z=z_h$, we fix the period of angle $\theta$ as $\beta$ (\ref{sect2: angle period}). Thus, for $T>0$, the left and right vertical dotted lines of  Fig. \ref{holo strip} are identified due to the periodicity of angle $\theta$. On the other hand, for the strip I with $T\le 0$, the `horizon' and potential conical singularity are hidden behind the branes (red/blue curves). Thus, the conical singularity is irrelevant, and $\theta$ can be non-periodic for $T\le 0$ \cite{Tadashi}.

\begin{figure}[t]
\centering
\includegraphics[width=10cm]{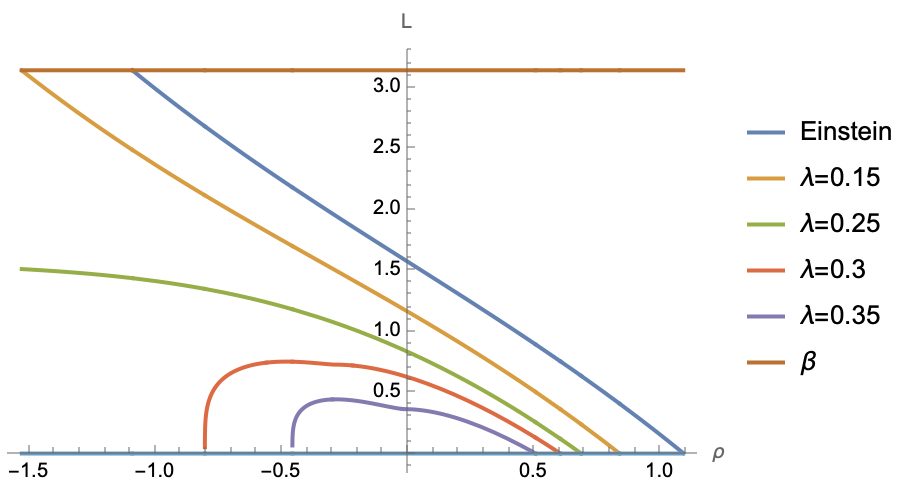} 
\caption{ Strip widths in normal phase with $0\le \lambda< 1/4$ and singular phase $1/4<\lambda\le 1/2$ for $d=4$.  Note that $L$ can be larger than $\beta$ for negative enough $\rho$ in the normal phase. We show only range of $L\le \beta$ for simplicity. On the other hand, $L$ is always smaller than $\beta$ in the singular phase. In both cases, the condition $0\le L$ imposes an upper bound of $\rho$. }
\label{strip width L4d}
\end{figure}

Below, we focus on the left half of branes with $T\le 0$, the right half can be obtained by symmetry.  Our convention of extrinsic curvature is $K=\nabla_{\mu} n^{\mu}$ with $n$ the outward-pointing normal vector. On the left halves of the EOW branes, we have
\begin{eqnarray}\label{sect2: nL}
n_{\mu}=\Big(\frac{S'(z)}{\sqrt{\frac{z^2 \left(h(z)^2 S'(z)^2+1\right)}{h(z)}}},\frac{-1}{\sqrt{\frac{z^2 \left(h(z)^2 S'(z)^2+1\right)}{f(z)}}},0,...,0\Big).
\end{eqnarray}
Substituting (\ref{sect2: AdS soliton}), (\ref{sect2: soliton Q}) and (\ref{sect2: nL}) into NBC (\ref{sect2: NBC}), we get an independent equation for $T\le 0$
\begin{eqnarray}\label{sect2: NBC solution}
T=-\frac{(d-1) h(z) \left((d-2) \lambda +h(z) S'(z) \sqrt{h(z) S'(z)^2+\frac{1}{h(z)}}\right)}{h(z)^2 S'(z)^2+1}. 
\end{eqnarray}
Let us verify that it corresponds to the left half of branes. From (\ref{sect2: Tension},\ref{sect2: NBC solution}), we derive $S'(0)=-\sinh(\rho)$ near $\theta_1$ of Fig. \ref{holo strip}, which agrees with (\ref{sect2: QL}) for the left boundary. As shown in Fig. \ref{holo strip}, we have $S'(z_{\text{max}})=\infty$ and $S'(z_{c})=0$ at the turning points $z_{\text{max}}$ and $z_c$ for the left halves of branes, respectively. Substituting $S'(z_{\text{max}})=\infty$ and $S'(z_{c})=0$ into (\ref{sect2: NBC solution}), we obtain 
\begin{eqnarray}\label{sect2: turning points}
T=-(d-1)\sqrt{h(z_{\text{max}})}=-(d-1)(d-2)\lambda h(z_c). 
\end{eqnarray}
Note that the turning point $z_c$ appears only for the red curve of Fig. \ref{holo strip} with $\rho>0$. Solving (\ref{sect2: NBC solution}) and (\ref{sect2: turning points}), we get 
\begin{eqnarray}\label{sect2: dS}
S'(z)=\pm \frac{\sqrt{2} \left(\sqrt{h\left(z_{\max }\right)}-(d-2) \lambda  h(z)\right)}{h(z) \sqrt{2 (d-2) \lambda  h(z) \sqrt{h\left(z_{\max }\right)}+h(z) H(z)-2 h\left(z_{\max }\right)+h(z)}},
\end{eqnarray}
where $H(z)=\sqrt{1+4 (d-2) \lambda  \left((d-2) \lambda  h(z)-\sqrt{h\left(z_{\max }\right)}\right)}$. 
We choose the positive sign, i.e., $S'(z)\sim \Big(\sqrt{h\left(z_{\max }\right)}-(d-2)\lambda  h(z)\Big) $, for the left halves of red and blue curves to get the correct behavior $S'(0)=-\sinh(\rho)$ near $\theta_1$ of Fig. \ref{holo strip}. The negative sign of (\ref{sect2: dS}) corresponds to the right halves of branes. From (\ref{sect2: soliton Q},\ref{sect2: dS}), we derive the width of strip I for $T\le 0$
\begin{eqnarray}\label{sect2: L I}
L_{\text{I}}=\int_0^{z_{\max}} dz\frac{2\sqrt{2} \left(\sqrt{h\left(z_{\max }\right)}-(d-2) \lambda  h(z)\right)}{h(z) \sqrt{2 (d-2) \lambda  h(z) \sqrt{h\left(z_{\max }\right)}+h(z) H(z)-2 h\left(z_{\max }\right)+h(z)}},
\end{eqnarray}
where $z_{\max}=(1-T^2/(d-1)^2)^{1/d}$ from (\ref{sect2: turning points}).  We stress that the above formula works for both the cases of blue curve and red curve of Fig. \ref{holo strip}. In particular, the integrand $2S'(z)\sim \Big(\sqrt{h\left(z_{\max }\right)}-(d-2)\lambda  h(z)\Big)$ flips signs at turning point $z_c$ (\ref{sect2: turning points}), which is the expected feature for the red curve. It suggests the integrand cannnot be chosen as the absolute value $2|S'(z)|\sim \Big|\sqrt{h\left(z_{\max }\right)}-(d-2)\lambda  h(z)\Big|$, which misses the case of red curve.  Recall that strip II is the complement of strip I and $(T, \lambda)$ of NBC (\ref{sect2: NBC}) flip signs when crossing the brane, we get the width of strip II for $T\ge 0$
\begin{eqnarray}\label{sect2: L II}
L_{\text{II}}=\beta - L_{\text{I}}\Big(T\to -T, \lambda\to -\lambda\Big).
\end{eqnarray}
In total, we have 
\begin{equation}\label{sect2: L}
L=\begin{cases}
\ L_{\text{I}},&\
\text{for} \ T\le 0,\\
\  L_{\text{II}},&\
\text{for} \ T\ge 0.
\end{cases}
\end{equation}
One can check $L$ is continuous at $T=0$, which is a test of our results. See Fig. \ref{strip width L4d} for example. 

Some comments are in order. First, (\ref{sect2: L I}) implies $L_{\text{I}}$ could be negative ($L_{\text{II}}=\beta-L_{\text{I}}$ could be larger than $\beta$ and result in conical singularity) for sufficiently large $\lambda$. To remove this unphysical case, we get an upper bound of the DGP coupling 
\begin{eqnarray}\label{sect2: range of lambda}
\lambda\le \frac{1}{d-2}.
\end{eqnarray}
It can be derived by considering the limit $z_{\max}\to 0$ ($T\to -(d-1)$). Then, $S'(z)\sim \Big(\sqrt{h\left(z_{\max }\right)}-(d-2)\lambda  h(z)\Big)\sim  \Big(1-(d-2)\lambda \Big)\ge 0 $ yields (\ref{sect2: range of lambda}).  See also Fig. \ref{strip width L4d}, which implies $L\to 0$ as $\lambda \to 1/(d-2)$. Second, the case $0 \le \lambda< 1/2(d-2)$ can continuously transform into that of Einstein gravity, while the case $1/2(d-2) < \lambda\le 1/(d-2)$ cannot. See Fig. \ref{strip width L4d} for instance. We name them normal and singular phases, respectively. Third, Fig. \ref{strip width L4d} shows the strip width $L$ decreases with brane tension $\rho$ in normal phase and $L\ge 0$ sets an upper bound of $\rho$ \cite{Miyaji:2021ktr} for fixed DGP parameter $\lambda$. For $T\ge 0$, $L$ should be smaller than the angle period $\beta$ to avoid the conical singularity. On the other hand, there is no upper bound of $L$ for $T<0$ since the conical singularity is hidden behind the brane \cite{Tadashi}. Thus, it is irrelevant to the bulk dual of the strip. As a result, there is no extra constraint on the lower bound of $T$, and it can take the minimal value $T=-(d-1)$ ($\rho\to -\infty$) \cite{Miyaji:2021ktr, Tadashi}.    

\begin{figure}[t]
\centering
\includegraphics[width=10cm]{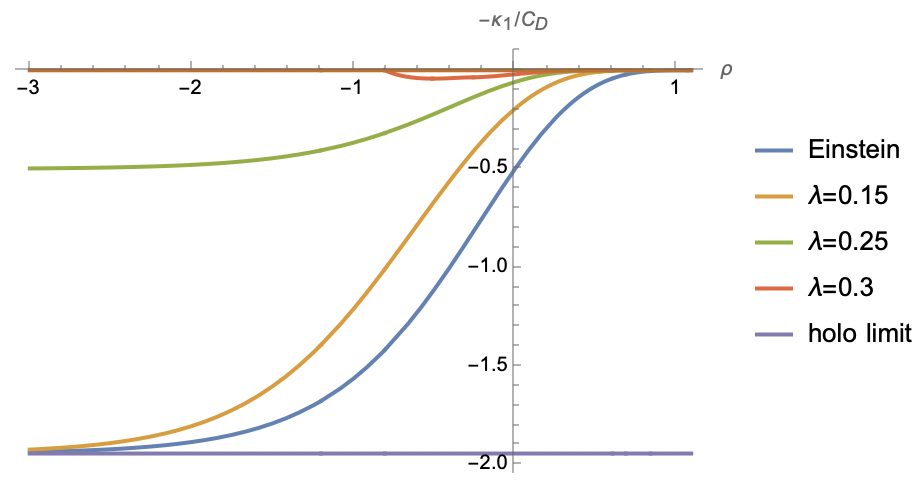} 
\caption{ $(-\kappa_1/C_D)$ in normal phase $0\le \lambda\le 1/4$ and singular phase with $1/4<\lambda\le1/2$ for $d=4$. In the normal phase, 
all curves approach the holographic limit (purple curve) from the above for $\rho\to -\infty$ ($T\to -3$). In the singular phase, $(-\kappa_1/C_D)$ is larger than the holographic limit (purple curve) and approach zero as $\lambda \to 1/2$. }
\label{bound1}
\end{figure}

We are prepared to analyze the $(-\kappa_1/C_D)$ ratio. The value of $\kappa_1$ can be determined from (\ref{sect2: kappa1}) with $z_h=1$ and (\ref{sect2: L I},\ref{sect2: L II},\ref{sect2: L}). The quantity $C_D$ is given by (\ref{sect2: CD d},\ref{sect2: CD 3d 4d}). We present the graph of $(-\kappa_1/C_D)$ in Fig. \ref{bound1}. In the normal phase with $0 \le \lambda< 1/2(d-2)$, the ratio $(-\kappa_1/C_D)$ increases with the tension $\rho$, reaching its minimum as $\rho\to -\infty$ ($T\to -(d-1)$). Notably, all curves converge to the same lower bound (purple curve) from above in the limit $\rho\to -\infty$ ($T\to -(d-1)$). We will provide an analytical proof of this observation in the following subsection. In the singular phase, $(-\kappa_1/C_D)$ is larger than the holographic limit (purple curve) and approaches zero as $\lambda \to 1/(d-2)$. Since we are interested in the lower bound of ($-\kappa_1/C_D$), we focus on the normal phase below.

\subsection{Analytical results}

Let us have some analytical discussions. We are interested in the case with minimal tension $T\to -(d-1)$ in the normal phase since it gives the holographic lower bound of the Casimir effect. Performing coordinate transformation $z=z_{\text{max}} y$ and expanding (\ref{sect2: L I}) around $x=\text{sech}^2(\rho)\to 0$ ($T\to -(d-1)$), we derive perturbatively in the normal phase
\begin{eqnarray}\label{sect2.2: L I}
L_{\text{I}}&=&\int_0^1 dy \ x^{1/d} \left(\frac{2 (1-2 (d-2) \lambda )^{\frac{1}{d}}}{\sqrt{1-y^d} \sqrt{x}}+O\left(\sqrt{x}\right)\right) \nonumber\\
&=&\frac{2x^{\frac{1}{d}-\frac{1}{2}}  \sqrt{\pi } (1-2 (d-2) \lambda )^{1/d} \Gamma \left(1+\frac{1}{d}\right)}{\Gamma \left(\frac{1}{2}+\frac{1}{d}\right)}+O\left(x^{\frac{1}{d}+\frac{1}{2}}\right).
\end{eqnarray}
For $x\to 0$ ($T\to -(d-1)$), the displacement operator (\ref{sect2: CD d}) becomes 
\begin{eqnarray}\label{sect2.2: CD}
C_D=\frac{2 x^{1-\frac{d}{2}} \left((d-1) \pi ^{\frac{1}{2}-\frac{d}{2}} (1-2 (d-2) \lambda ) \Gamma (d+2)\right)}{d \Gamma \left(\frac{d+1}{2}\right)}+O(x^{2-\frac{d}{2}} ).
\end{eqnarray}
In the limit $x\to 0$ ($T\to -(d-1)$), we obtain the universal lower bound (\ref{Casimir bound}) independent of $\lambda$
\begin{eqnarray}\label{sect2.2: limit}
\lim_{T\to -(d-1)} (-\frac{\kappa_1}{C_D})=\lim_{x\to 0} (-\frac{L_{\text{I}}^d}{C_D})=-\frac{2^{d-2} d \pi ^{d-\frac{1}{2}} \Gamma \left(\frac{d-1}{2}\right) \Gamma \left(\frac{1}{d}\right)^d}{\Gamma (d+2) \left(d \Gamma \left(\frac{1}{2}+\frac{1}{d}\right)\right)^d},
\end{eqnarray}
in the normal phase with $0 \le \lambda< 1/2(d-2)$. Interestingly, unlike the KSS bound, holography imposes a universal lower bound of the Casimir effect. The following sections show this is also the case for GB gravity and GB-DGP gravity.  

We need the negative brane tension to derive the holographic lower bound of the Casimir effect (\ref{sect2.2: limit}). Unlike the case of two-side branes in the discontinuous spacetime with non-trivial junction condition, the negative brane tension is well-defined on the one-side end-of-the-world (EOW) branes in AdS/BCFT \cite{Takayanagi:2011zk, Fujita:2011fp, Miyaji:2021ktr, Tadashi}. Recall that the brane tension flips signs when crossing the EOW brane in the continuous spacetime \footnote{For the continuous spacetime, we have trivial junction condition $K_{L\ ij}+K_{R\ ij}=0$ with $n_L=-n_R$ on the two sides of the brane.}. If the positive brane tension is stable from the viewpoint of one side (gravity dual of strip II), so is the negative brane tension from the other side's perspective (gravity dual of strip I). Besides, the brane tension is a cosmological constant rather than a kinetic-energy term on the EOW brane. Of course, the negative cosmological constant is well-defined in AdS/CFT. One can prove the gravitational KK modes are ghost-free and tachyon-free on the EOW brane with general tensions (including negative brane tension) \cite{Miao:2023mui}. See Appendix A for an example. Finally, the negative brane tension would yield a negative A-type boundary central charge. But nothing goes wrong. Recall that the A-type boundary central charge is negative for a free scalar with the Dirichlet boundary condition \cite{Jensen:2015swa}. For the reasons above, it is well-defined for the AdS/BCFT with a negative brane tension. 

\section{Holography II: GB gravity}

This section explores the holographic bound of the Casimir effect in Gauss-Bonnet (GB) gravity. Since the calculations are similar to those in DGP gravity, we will present only the key points below. We take the following form of GB action \cite{Hu:2022ymx} \footnote{Note that (\ref{sect3: IGBmiao}) is not the usual action of GB gravity. The relation to the usual one \cite{Buchel:2009sk} can be found in \cite{Hu:2022ymx}. We take the action (\ref{sect3: IGBmiao}) because it simplifies the calculations of displacement operator $C_D$. Besides, the AdS radius can be set to one $l=1$ as Einstein gravity  \cite{Hu:2022ymx}.}
 \begin{eqnarray}\label{sect3: IGBmiao}
&&I_{\text{GB}}=\int_N d^{d+1}x\sqrt{|g|} \Big(R+d(d-1)+\alpha \ \mathcal{L}_{\text{GB}}(\bar{R}) \Big)\nonumber\\
&&\ \ \ \ \ \ + 2\int_Q d^{d}y\sqrt{|h|} \Big( (1+2(d-1)(d-2)\alpha)(K-T)+ 2\alpha  (J-2 G^{ij}_{Q} K_{ij}) \Big),
 \end{eqnarray}
 where $\mathcal{L}_{\text{GB}}(\bar{R}) =\bar{R}_{\mu\nu\alpha\beta}\bar{R}^{\mu\nu\alpha\beta}-4\bar{R}_{\mu\nu}\bar{R}^{\mu\nu}+\bar{R}^2$,  $\bar{R}$ is defined as
\begin{eqnarray}\label{sect3: background curvature1}
&&\bar{R}=R+d(d+1),\\ \label{background curvature2}
&&\bar{R}_{\mu\nu}=R_{\mu\nu}+d g_{\mu\nu},\\ \label{background curvature3}
&&\bar{R}_{\mu\nu\rho\sigma}=R_{\mu\nu\rho\sigma}+(g_{\mu\rho}g_{\nu\sigma}-g_{\mu\sigma}g_{\nu\rho}),
\end{eqnarray}
which vanish on AdS space with unite radius. $G^{ij}_{Q} $ is the intrinsic Einstein tensor on the boundary $Q$, and  $J$ is the trace of 
 \begin{eqnarray}\label{sect3: Jij}
J_{ij}=\frac{1}{3}\left(2 K K_{ik}K^k_j-2 K_{ik}K^{kl}K_{lj}+K_{ij}\left(K_{kl}K^{kl}-K^2\right) \right). 
  \end{eqnarray}
 From action (\ref{sect3: IGBmiao}), we derive NBC on the brane $Q$
 \begin{eqnarray}\label{sect3: NBC}
\left(1+2(d-1)(d-2)\alpha \right)\Big(K^{ij}-\left(K-T\right)h^{ij}\Big) +2\alpha (H^{ij}-\frac{1}{3}H h^{ij})=0,
\end{eqnarray}
where    
\begin{eqnarray}\label{sect3: Qij}
H_{ij}=3J_{ij}+2 K \mathcal{R}_{ij}+\mathcal{R} K_{ij}-2K^{kl} \mathcal{R}_{kilj}-4\mathcal{R}_{k(i}K^k_{j)}.
 \end{eqnarray}
 Following the same logic of sect. 2, we parameterize the brane tension as \cite{Hu:2022ymx} 
  \begin{eqnarray}\label{sect3: tension}
T=\frac{(d-1) \tanh (\rho ) \text{sech}^2(\rho ) ((4 \alpha  (d-2) d+3) \cosh (2 \rho )-4 \alpha  (d-6) (d-2)+3)}{6+12 \alpha  (d-2) (d-1)}.
 \end{eqnarray} 
  To avoid negative energy fluxes, the GB coupling $\alpha$ should obey \cite{Hu:2022ymx,Buchel:2009sk}
   \begin{eqnarray}\label{sect3: GBconstraintbar}
   \frac{-1}{4(d^2-2d-2)}\le \alpha \le \frac{1}{8}.
  \end{eqnarray} 
 By applying the method of \cite{Miao:2023mui}, we derive the ghost-free condition for gravitational KK modes of GB gravity (\ref{sect3: IGBmiao})
\begin{eqnarray}\label{sect3: ghost-free condition}
 \frac{ \alpha  \tanh (\rho ) }{1+4 \alpha(d-2)} \ge 0.
\end{eqnarray} 
See appendix A for the derivation of (\ref{sect3: ghost-free condition}) with $d=4$. 
The norm of displacement operator is given by \cite{Hu:2022ymx} for $\rho\le 0$
\begin{eqnarray}\label{sect3: CD}
C_D=-\frac{4 \pi ^{\frac{1}{2}-\frac{d}{2}} (4 \alpha  (d-2)+1) \Gamma (d+2)}{\Gamma \left(\frac{d-1}{2}\right)} \frac{\cosh ^d(\rho ) \left((4 \alpha  (d-2)+1) \coth ^2(\rho )+4 \alpha  (d-3) (d-2)\right)}{d (4 \alpha  (d-2)+1) \cosh ^2(\rho ) \coth ^3(\rho )+G (-\coth (\rho ))^d},
\end{eqnarray} 
where 
\begin{eqnarray}\label{sect3: GBG}
G=\left((4 \alpha  (d-2)+1) \coth ^2(\rho )+4 \alpha  (d-3) (d-2)\right) \, _2F_1\left(\frac{d-1}{2},\frac{d}{2};\frac{d+2}{2};-\text{csch}^2(\rho )\right).
\end{eqnarray}
Similar to DGP gravity, we can make analytical continuation to get $C_D$ with $\rho\ge 0$. For $d=4$, we get
\begin{eqnarray}\label{sect3: CD 4d}
(C_D)_{4d}=\frac{120 (8 \alpha +1) \coth (\rho ) ((16 \alpha +1) \cosh (2 \rho )+1)}{\pi ^2 (\coth (\rho )+1) (4 \alpha  \sinh (2 \rho )+(12 \alpha +1) \cosh (2 \rho )+4 \alpha +1)}.
\end{eqnarray} 

The gravity dual of the vacuum of the strip is given by AdS soliton (\ref{sect2: AdS soliton}) with
\begin{eqnarray}\label{sect3: f 4d}
h(z)=\frac{1+2 \alpha  (d-1) (d-2)-\sqrt{(1+4 \alpha  (d-2))^2+4 \alpha  (d-3) (d-2) \left(\alpha  \left(d^2-d-2\right)+1\right)\frac{ z^d}{z_h^d}}}{2 \alpha  (d-2) (d-3)},
\end{eqnarray}
where $h(z_h)=0$. Substituting the embedding function $\theta=S(z)$ and metric (\ref{sect2: AdS soliton}) into the NBC (\ref{sect3: NBC}), we obtain one independent equation for the left halves of branes with $T\le 0$
 \begin{eqnarray}\label{sect3: key equation1}
T=\frac{(1-d) h(z) S'(z)}{\sqrt{h(z) S'(z)^2+\frac{1}{h(z)}}}+\frac{2 \alpha  (d-1) (d-2) (d-3) h(z)^2 S'(z) \left(h(z)^2 S'(z)^2+3\right)}{3 (1+2 \alpha  (d-1) (d-2)) \sqrt{h(z) S'(z)^2+\frac{1}{h(z)}} \left(h(z)^2 S'(z)^2+1\right)}.
\end{eqnarray}
Substituting the critical points $S'(z_{\max})=\infty$ into the above equation, we get
 \begin{eqnarray}\label{sect3: key equation2}
T= \frac{(d-1) \sqrt{h\left(z_{\max }\right)} \left(-6 \alpha  \left(d^2-3 d+2\right)+2 \alpha  \left(d^2-5 d+6\right) h\left(z_{\max }\right)-3\right)}{6 \alpha  \left(d^2-3 d+2\right)+3}.
\end{eqnarray}
From (\ref{sect3: key equation1}) and (\ref{sect3: key equation2}), we can solve $S'(z)$ in functions of $h(z)$ and $h(z_{\max})$. There are multiple solutions for  $S'(z)$, and we choose the one that takes real values and can yield the correct limit $S'(0)=-\sinh(\rho)$ near $\theta_1$ of Fig. \ref{holo strip}. We do not show the complicated expression of $S'(z)$ for simplicity.  From $S'(z)$, we can obtain the width for strip
\begin{equation}\label{sect3: L}
L=\begin{cases}
\ L_{\text{I}}=\int_0^{z_{\max}} 2S'(z)dz,&\
\text{for} \ T\le 0,\\
\  L_{\text{II}}=\beta-L_{\text{I}}(\rho\to -\rho, \alpha\to \alpha),&\
\text{for} \ T\ge 0.
\end{cases}
\end{equation}
Note that the GB coupling $\alpha$ has different sign transformation rule from $\rho$ in $L_{
\text{II}}$. That is because $\alpha$ appears in terms like $K \mathcal{R}^2$ and $K^3$ on the brane, which flip signs when crossing the brane (similar to $K$). On the other hand,  the tension $T$ keeps invariant when crossing the brane. To keep the form of NBC (\ref{sect3: NBC}) invariant, the tension $T$ should flip signs while GB coupling $\alpha$ remains invariant when crossing the brane. By using the holographic renormalization for GB gravity \cite{Sen:2014nfa}, we obtain the energy density $T_{tt}=-(1+\alpha  (d-2) (d+1))/z_h^d$. Recall that $T_{tt}=-\kappa_1/L^d$ for a strip. Then, we get the holographic Casimir amplitude
 \begin{eqnarray}\label{sect3: Casimir coefficient for GB}
\kappa_1=\Big(1+\alpha  (d-2) (d+1)\Big) \frac{L^d}{z_h^d}. 
\end{eqnarray}

 \begin{figure}[t]
\centering
\includegraphics[width=10cm]{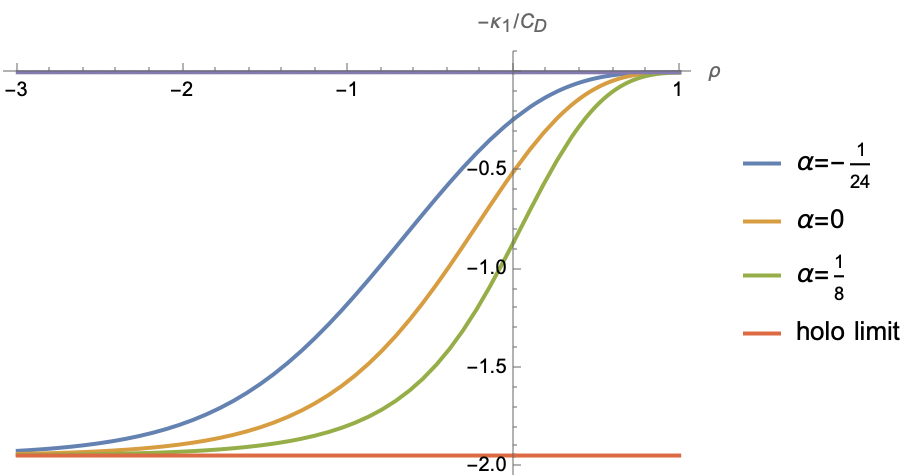}
\caption{ $(-\kappa_1/C_D)$ for GB gravity with $d=4$. The blue, orange, green, red curves denote GB gravity with typical couplings $\alpha=(-\frac{1}{24}, 0,\frac{1}{8})$, and holographic limit (\ref{Casimir bound}), respectively. It shows GB gravity with various couplings $\alpha$ approach the same value in the limit $\rho\to -\infty$. }
\label{bound of 4d GB}
\end{figure}

Now we are ready to discuss the holographic lower bound of $(-\kappa_1/C_D)$. For simplicity, we set $z_h=1$ below. We draw $(-\kappa_1/C_D)$ for GB gravity with typical couplings (\ref{sect3: GBconstraintbar}) for $d=4$ in Fig. \ref{bound of 4d GB}. Unlike DGP gravity, there is no singular phase for the GB gravity with couplings obeying (\ref{sect3: GBconstraintbar}). Like DGP gravity in the normal phase, $(-\kappa_1/C_D)$ for GB gravity increases with $\rho$ and approaches the universal lower bound in the limit $\rho\to -\infty$. To end this section, we analytically derive the holographic lower bound of the Casimir effect for GB gravity. Fig. \ref{bound of 4d GB} shows the lower bound is saturated in the limit $x=\text{sech}^2(\rho)\to 0$ ($\rho\to -\infty$). In such limit, we can solve perturbatively $S'(z)$ from (\ref{sect3: key equation1}) and then derive the strip width (\ref{sect3: L})
 \begin{eqnarray}\label{sect3: L small x}
L_{\text{I}}&=&\int_0^1 dy \frac{2 \left(\frac{4 \alpha  (d-2)^2+1}{\alpha  \left(d^2-d-2\right)+1}\right)^{1/d} x^{\frac{1}{d}-\frac{1}{2}}}{\sqrt{1-y^d}}+O(x^{\frac{1}{d}+\frac{1}{2}})\nonumber\\
&=&\frac{2 \sqrt{\pi } \left(\frac{4 \alpha  (d-2)^2+1}{\alpha  \left(d^2-d-2\right)+1}\right)^{1/d} \Gamma \left(1+\frac{1}{d}\right)}{\Gamma \left(\frac{1}{2}+\frac{1}{d}\right)} x^{\frac{1}{d}-\frac{1}{2}}+O(x^{\frac{1}{d}+\frac{1}{2}}).
\end{eqnarray}
For $x=\text{sech}^2(\rho)\to 0$, the displacement operator (\ref{sect3: CD}) becomes 
 \begin{eqnarray}\label{sect3: CD small x}
C_D=\frac{x^{1-\frac{d}{2}} \left(2 (d-1) \pi ^{\frac{1}{2}-\frac{d}{2}} \left(4 \alpha  (d-2)^2+1\right) \Gamma (d+2)\right)}{d \Gamma \left(\frac{d+1}{2}\right)}+O(x^{2-\frac{d}{2}}).
\end{eqnarray}
From (\ref{sect3: Casimir coefficient for GB},\ref{sect3: L small x},\ref{sect3: CD small x}) with $z_h=1$, we finally derive the universal limit for GB gravity
\begin{eqnarray}\label{sect3: holo limit}
\lim_{x\to 0} (-\frac{\kappa_1}{C_D})=\lim_{x\to 0} (-\frac{\Big(1+\alpha  (d-2) (d+1)\Big) L_{\text{I}}^d}{C_D})=-\frac{2^{d-2} d \pi ^{d-\frac{1}{2}} \Gamma \left(\frac{d-1}{2}\right) \Gamma \left(\frac{1}{d}\right)^d}{\Gamma (d+2) \left(d \Gamma \left(\frac{1}{2}+\frac{1}{d}\right)\right)^d},
\end{eqnarray}
which verifies again that holography imposes a universal lower bound (\ref{Casimir bound}) for the Casimir effect.

\section{Holography III: GB-DGP gravity}  

This section studies the holographic bound of the Casimir effect in GB-DGP gravity. The qualitative behaviors of the Casimir effect in this context are similar to those observed in DGP and GB gravity. For simplicity, we focus on the dimension $d=4$ and show only the key points below. 

Let us quickly recall some key points. The action of GB-DGP gravity for $d=4$ reads
 \begin{eqnarray}\label{sect4: IGBmiao}
&&I_{\text{GB-DGP}}=\int_N d^{5}x\sqrt{|g|} \Big(R+12+\alpha \ \mathcal{L}_{\text{GB}}(\bar{R}) \Big)\nonumber\\
&&\ \ \ \ \ \ + 2\int_Q d^{4}y\sqrt{|h|} \Big( (1+12\alpha)(K-T+\lambda \mathcal{R})+ 2\alpha  (J-2 G^{ij}_{Q} K_{ij}) \Big),
 \end{eqnarray}
with NBC on the brane
 \begin{eqnarray}\label{sect4: NBC}
\left(1+12\alpha \right)\Big(K^{ij}-\left(K-T+\lambda \mathcal{R}\right)h^{ij}+2\lambda \mathcal{R}^{ij}\Big) +2\alpha (H^{ij}-\frac{1}{3}H h^{ij})=0,
\end{eqnarray}
where $H$ is given by (\ref{sect3: Qij}) and the brane tension is parameterized as 
 \begin{eqnarray}\label{sect4: tension}
T=\frac{\text{sech}^3(\rho ) (-24 (12 \alpha +1) \lambda  \cosh (\rho )+(32 \alpha +3) \sinh (3 \rho )+3 \sinh (\rho ))}{48 \alpha +4}.
 \end{eqnarray} 
 To avoid negative energy fluxes and to be ghost-free, the GB and DGP couplings should obey 
   \begin{eqnarray}\label{sect4: GBconstraintbar}
   \frac{-1}{24}\le \alpha \le \frac{1}{8}.
  \end{eqnarray} 
  and 
  \begin{eqnarray}\label{sect4: ghost-free condition}
 \frac{(1+12 \alpha)  \lambda +2 \alpha  \tanh (\rho ) }{1+8 \alpha} \ge 0. 
\end{eqnarray}
See Appendix A for the derivation of the ghost-free condition (\ref{sect4: ghost-free condition}). The norm of displacement operator reads 
\begin{eqnarray}\label{sect4: CD 4d}
C_D=\frac{240 (8 \alpha +1) e^{\rho } \cosh (\rho ) (4 (12 \alpha +1) \lambda  \sinh (2 \rho )+(16 \alpha +1) \cosh (2 \rho )+1)}{\pi ^2 \left(e^{4 \rho } (16 \alpha  (3 \lambda +1)+4 \lambda +1)+e^{2 \rho } (8 \alpha  (6 \lambda +1)+4 \lambda +2)+8 \alpha +1\right)},
\end{eqnarray} 
where the calculations are given by Appendix B. Substituting the brane embedding function $\theta=S(z)$ and AdS soliton metric (\ref{sect2: AdS soliton}) into NBC (\ref{sect4: NBC}), we get one independent equation for $T\le 0$
 \begin{eqnarray}\label{sect4: key equation1}
T=-\frac{h(z) \left(6 (12 \alpha +1) \lambda  \sqrt{h(z) S'(z)^2+\frac{1}{h(z)}}+h(z)^2 (36 \alpha -4 \alpha  h(z)+3) S'(z)^3+(36 \alpha -12 \alpha  h(z)+3) S'(z)\right)}{(12 \alpha +1) \sqrt{h(z) S'(z)^2+\frac{1}{h(z)}} \left(h(z)^2 S'(z)^2+1\right)}.
\end{eqnarray}
Substituting the critical points $S'(z_{\max})=\infty$ and  $S'(z_c)=0$ into the above equation, we get
 \begin{eqnarray}\label{sect4: key equation2}
T=\sqrt{h\left(z_{\max }\right)} \left(\frac{4 \alpha  h\left(z_{\max }\right)}{12 \alpha +1}-3\right)=-6 \lambda  h\left(z_c\right). 
\end{eqnarray}
From (\ref{sect4: key equation1}) and (\ref{sect4: key equation2}), we can express $S'(z)$ in terms of $f(z)$ and $f(z_{\max})$. There are multiple solutions for $S'(z)$, and we choose the real one with the correct limit $S'(0)=-\sinh(\rho)$ near $\theta_1$ of Fig. \ref{holo strip}. From $S'(z)$, we get the strip width
\begin{equation}\label{sect3: L}
L=\begin{cases}
\ L_{\text{I}}=\int_0^{z_{\max}} 2S'(z)dz,&\
\text{for} \ T\le 0,\\
\  L_{\text{II}}=\beta-L_{\text{I}}(\rho\to -\rho, \lambda\to -\lambda, \alpha\to \alpha),&\
\text{for} \ T\ge 0.
\end{cases}
\end{equation}

\begin{figure}[t]
\centering
\includegraphics[width=7.5cm]{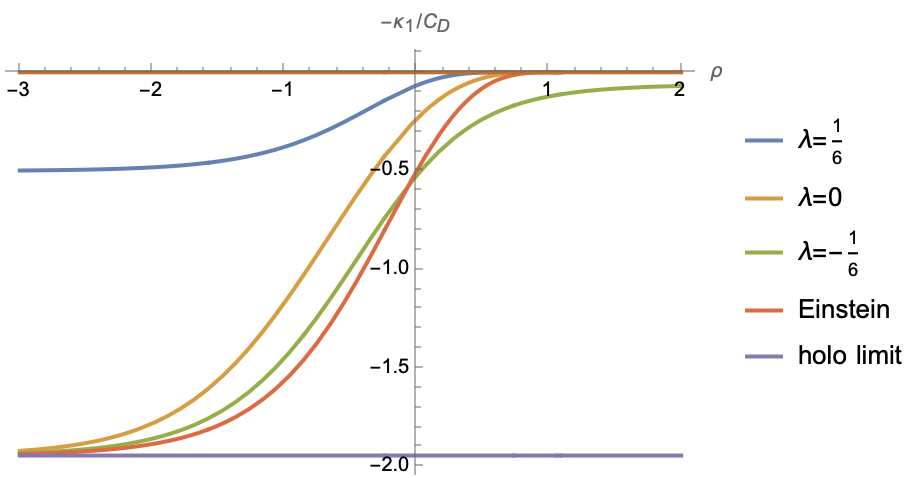} \includegraphics[width=7.5cm]{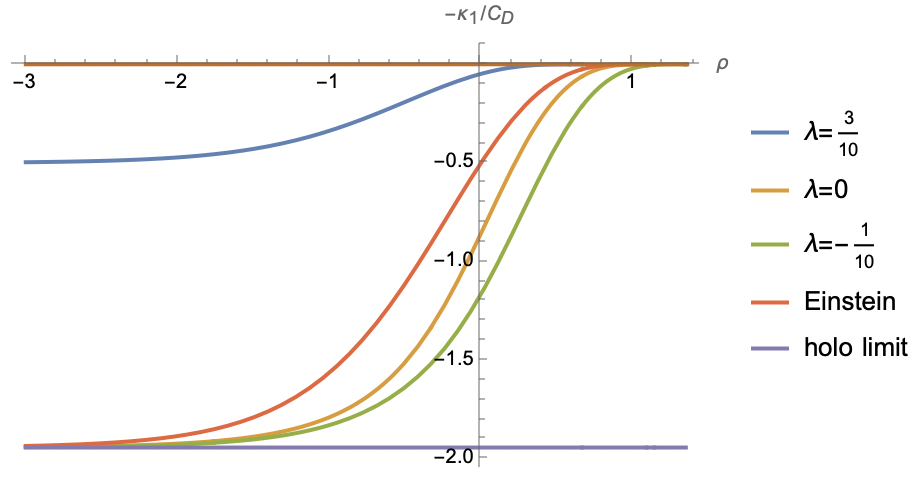}
\caption{ $(-\kappa_1/C_D)$ for GB-DGP gravity with $\alpha=-1/24$ (left) and $\alpha=1/8$ (right). We focus on normal phase (\ref{sect3.3: normal phase}) and $d=4$. The blue and green curves correspond to upper and lower bounds of  (\ref{sect3.3: normal phase}).  All curves except the blue ones approach the universal holographic limit (purple curve) for $\rho\to -\infty$. The blue curve corresponds to the critical point of normal and singular phases. For $\alpha=-1/24$ and $\lambda=-1/6$, $(-\kappa_1/C_D)$ (green curve) cannot approach zero. }
\label{bound of 4d GBDGP}
\end{figure}

Let us focus on the normal phase, which yields smaller $(-\kappa_1/C_D)$ than the singular phase. The parameter space for the normal phase is
 \begin{eqnarray}\label{sect3.3: normal phase}
-\frac{2 \alpha  \tanh (\rho )}{12 \alpha +1} \le \lambda <  \frac{1+16 \alpha}{4(1+12 \alpha )},
 \end{eqnarray}  
where the lower bound is derived from the ghost-free condition (\ref{sect4: ghost-free condition}) and the upper bound is obtained from $C_D\ge 0$ (\ref{sect4: CD 4d}) or the small $x$ expansion (\ref{sect4: LI expand}). Note that (\ref{sect3.3: normal phase}) reduces to the normal phase $0 \le \lambda<1/4$ for DGP gravity with $\alpha=0$, which is a test of our calculations. We draw $(-\kappa_1/C_D)$ for GB-DGP gravity in the normal phase in Fig. \ref{bound of 4d GBDGP}. It shows all curves except the blue ones approach the universal holographic limit (purple curve) for $\rho\to -\infty$. Note that the blue curve corresponds to the phase-transition point of normal and singular phases. It cannot saturate but still obeys the holographic lower bound (\ref{Casimir bound}). Following the method of sect.3.1, we consider the limit $x=\text{sech}^2(\rho)\to 0$ ($\rho\to -\infty$), which tends to give smaller $(-\kappa_1/C_D)$.  In such limit, we can solve perturbatively $S'(z)$ from (\ref{sect4: key equation1}) and then derive the strip width (\ref{sect3: L}) in the normal phase
 \begin{eqnarray}\label{sect4: LI expand}
L_{\text{I}}=2\sqrt{\pi } \frac{\Gamma \left(\frac{5}{4}\right)}{\Gamma \left(\frac{3}{4}\right)} (\frac{ 1-4 \lambda -48 \alpha  \lambda +16 \alpha}{(1+10 \alpha ) \ x})^\frac{1}{4}+O\left(x^{\frac{3}{4}}\right).
 \end{eqnarray}
 In the limit $x=\text{sech}^2(\rho)\to 0$, the norm of displacement operator (\ref{sect4: CD 4d}) becomes 
  \begin{eqnarray}\label{sect4: CD expand}
C_D=\frac{240 (1-4 \lambda -48 \alpha  \lambda +16 \alpha)}{\pi ^2 x}+O\left(x^0\right).
 \end{eqnarray}
 Then, we obtain the universal limit for the GB-DGP gravity in the normal phase
  \begin{eqnarray}\label{sect4: holo limit}
\lim_{x\to 0}(\frac{-\kappa_1}{C_D})=\lim_{x\to 0}\frac{-(1+10\alpha) L_{\text{I}}^4}{C_D}=-\frac{\pi ^4 \Gamma \left(\frac{5}{4}\right)^4}{15 \Gamma \left(\frac{3}{4}\right)^4},
 \end{eqnarray}
 which reproduces the lower bound (\ref{Casimir bound}) of Casimir effect for $d=4$.

\section{Tests of holographic bound}

This section tests the Casimir effect's lower bound  (\ref{Casimir bound}). We verify it by free scalars and fermions in general dimensions, Maxwell theory for $d=4$ and $O(N)$ models in the $\epsilon=4-d$ expansions.  

\begin{figure}[t]
\centering
\includegraphics[width=10cm]{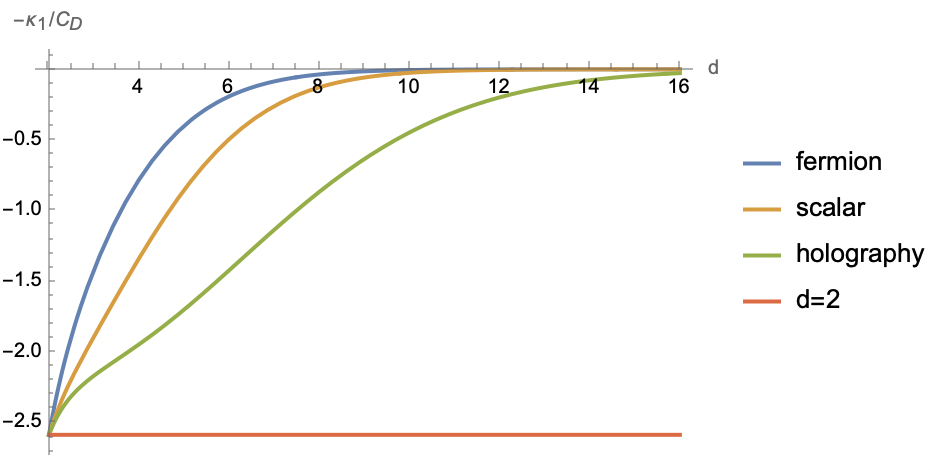} 
\caption{Various ratios $(-\kappa_1/C_D)$ in general dimensions. It shows $(-\kappa_1/C_D)_{\text{fermion}}\ge (-\kappa_1/C_D)_{\text{scalar}}\ge (-\kappa_1/C_D)_{\text{holo}}$. Besides, all ratios coincide for $d=2$ and approach zero for $d\to \infty$. }
\label{ratios HD}
\end{figure}

We first study free BCFTs. For free scalar with Robin boundary condition (RBC) and Dirichlet boundary condition (DBC), the Casimir amplitude \cite{Romeo:2000wt} and displacement operator \cite{Miao:2018dvm} are given by
  \begin{eqnarray}\label{sect5: kappa1 scalar}
&&(\kappa_1)_s=\frac{\zeta (d) \Gamma \left(\frac{d}{2}\right)}{(4 \pi )^{d/2}},\\
&&(C_D)_s=\frac{\Gamma \left(\frac{d}{2}\right)^2}{2 \pi ^d}, \label{sect5: CD scalar}
 \end{eqnarray}
 which yields the ratio
   \begin{eqnarray}\label{sect5: ratio scalar}
(\frac{-\kappa_1}{C_D})_s=-\frac{2^{1-d} \pi ^{d/2} \zeta (d)}{\Gamma \left(\frac{d}{2}\right)}.
 \end{eqnarray}
 For massless Dirac fermions, the Casimir amplitude \cite{Bellucci:2009hh} and displacement operator \cite{McAvity:1993ue} \footnote{In the notation of \cite{McAvity:1993ue}, we have $C_D=\alpha(1)=-(d-1)\beta(1)$. Note that we have replaced $2^{\frac{d}{2}}$ of \cite{McAvity:1993ue} with $2^{[\frac{d}{2}]}$, since we adopt $N\times N$ gamma matrices with $N=2^{[\frac{d}{2}]}$.} read 
   \begin{eqnarray}\label{sect5: kappa1 fermion}
&&(\kappa_1)_f= \frac{\zeta (d) \Gamma \left(\frac{d}{2}\right)\left(1-2^{-(d-1)}\right) 2^{\left[\frac{d}{2}\right]}}{(4 \pi )^{d/2}},\\
&&(C_D)_f=(d-1) \pi ^{-d} \Gamma \left(\frac{d}{2}\right)^2 2^{\left[\frac{d}{2}\right]-2}, \label{sect5: CD fermion}
  \end{eqnarray}
which gives 
   \begin{eqnarray}\label{sect5: ratio fermion}
(\frac{-\kappa_1}{C_D})_f=-\frac{4^{1-d} \left(2^d-2\right) \pi ^{d/2} \zeta (d)}{(d-1) \Gamma \left(\frac{d}{2}\right)}.
 \end{eqnarray}
We draw various ratios $(-\kappa_1/C_D)$ in Fig. \ref{ratios HD}, which shows $(-\kappa_1/C_D)_f\ge (-\kappa_1/C_D)_s\ge (-\kappa_1/C_D)_{\text{holo}}$ in general dimensions. For $d=2$, all ratios coincide
   \begin{eqnarray}\label{sect5: ratio 2d}
(\frac{-\kappa_1}{C_D})|_{d=2}=-\frac{\pi ^3}{12}\approx -2.58,
 \end{eqnarray}
 which is the expected result for 2d BCFTs \footnote{For $d=2$, the strip Casimir effect can be obtained from that of half space by the conformal map. As a result, we have $\kappa_1=c \pi/24$ \cite{Bloete:1986qm} and $C_D=c/(2\pi^2)$. Thus, we have universally $(-\kappa_1/C_D)=-\pi^3/12$ for 2d BCFTs. Note that we impose the same conformal boundary conditions on the two boundaries of the strip. }. Besides, all ratios approach zero as $d\to \infty$. 

Let us go on to discuss Maxwell's theory for $d=4$. We have 
   \begin{eqnarray}\label{sect5: kappa1 CD Maxwell}
(\kappa_1)_M=\frac{\pi^2}{720 }, \ \  (C_D)_M=\frac{6}{\pi^4}, 
  \end{eqnarray}
 which yields 
    \begin{eqnarray}\label{sect5: ratio 2d}
(\frac{-\kappa_1}{C_D})|_{M}=-\frac{\pi ^6}{4320}\approx -0.22.
 \end{eqnarray}
 We summarize various free BCFTs for $d=3$ and $d=4$ in Table. \ref{table1 Casimir bounds}, which all obeys the holographic lower bound (\ref{Casimir bound}) of Casimir effect.  
 
We focus on the free BCFTs above. Now, let us consider BCFTs with interactions. We remark that the 3d Ising model and $O(N)$ model with $N=2,3$ obey the lower bound (\ref{Casimir bound}) of the Casimir effect \cite{Miao:2024gcq}. Consider $O(N)$ in the $\epsilon=4-d$ expansions. The Casimir amplitude \cite{Krech:Casimir Effect} and the norm of Casimir displacement operator \cite{McAvity:1993ue} are given by
   \begin{eqnarray}\label{sect5: kappa1 ON}
&&(\kappa_1)_{O(N)}=\frac{\pi ^2 N}{1440}\left( 1+ \epsilon  \left(\frac{\gamma -1}{2}-\frac{5 (N+2)}{4 (N+8)}-\frac{\zeta '(4)}{\zeta (4)}+\log \left(2 \sqrt{\pi }\right)\right) \right),\\
&&(C_D)_{O(N)}=\frac{N}{2 \pi ^4} \left(1+\epsilon  \left(\gamma \pm\frac{5   (N+2)}{6 (N+8)}+\log (\pi )-1\right)\right),\label{sect5: CD ON}
  \end{eqnarray}
 where $\pm$ denotes RBC and DBC. We get the ratio
\begin{eqnarray}\label{sect5: ratio ON}
(\frac{-\kappa_1}{C_D})|_{O(N)}=\frac{-\pi ^6}{720}\left(1+ \epsilon  \left(\frac{1-\gamma }{2}-\frac{90 \zeta '(4)}{\pi ^4}\mp \frac{5   (N+2)}{6 (N+8)}-\frac{5 (N+2)}{4 (N+8)}+\log \left(\frac{2}{\sqrt{\pi }}\right)\right)\right),
\end{eqnarray}
where $\gamma\approx 0.58$ is Euler's constant. Note that the ratio (\ref{sect5: ratio ON}) is larger than that of free scalar (\ref{sect5: ratio scalar}) with $d=4-\epsilon$
\begin{eqnarray}\label{sect5: ratio difference}
(\frac{-\kappa_1}{C_D})|_{O(N)}-(\frac{-\kappa_1}{C_D})|_{s}\approx  \frac{\pi ^6 (N+2)}{1728 (N+8)}(3\pm 2) \epsilon>0.
\end{eqnarray}
Thus, it obeys the holographic lower bound (\ref{Casimir bound}). Here we require $\epsilon=g (N+8)/48 \pi ^2>0$ to avoid negative $\phi^4$ interaction. Usually, a bound is set either by free theories or strongly coupled CFTs dual to gravity. For example, the Bueno-Casini-Andino-Moreno (BCAM) bound for entanglement entropy is set by free scalar \cite{Bueno:2023gey}, 
while the KSS bound \cite{Kovtun:2004de} for the fluid is set by holography \cite{Kovtun:2004de}. The $O(N)$ in the $\epsilon=4-d$ expansion suggests that the free scalar may set the lower bound of the Casimir effect. However, the $O(N)$ models in non-integer dimensions are non-unitary \cite{Hogervorst:2015akt}. On the other hand, the Ising with $d=3$ confirms that holography, instead of the free scalar, sets the lower bound of the Casimir effect \cite{Miao:2024gcq, Toldin:2021kun, Ising referee1, Ising referee2}. The BCAM bound conjectures \cite{Bueno:2023gey}
\begin{eqnarray}\label{sect5: BCAM bound}
\frac{C_T}{F_0}\le \frac{C_T}{F_0}|_{\text{free scalar}},
\end{eqnarray}
where $C_T$ is the stress-tensor two-point function coefficient for CFT, corresponding to $C_D$ for BCFT, and $F_0$ is the entanglement entropy universal coefficient across a round spherical entangling surface, obeying F-theorem \cite{Myers:2010tj}. The Casimir amplitude $\kappa_1$ differs from $F_0$, which does not obey the F-theorem or g-theorem. Thus, it is unsurprising that $C_T/F_0$ and  $C_D/\kappa_1$ satisfy different bounds.

\section{Holographic bound for wedge}

This section investigates the holographic bound of the Casimir effect for a wedge, which is the simplest generalization of a strip. For simplicity, we focus on Einstein's gravity and leave the discussions of higher derivative gravity to future work. Recall the wedge Casimir effect takes the form (\ref{Tij wedge}) in a ground state. The wedge Casimir amplitude $f(\Omega)$ has two interesting limits
\begin{eqnarray}\label{sect6: limits}
f(\Omega)\to \begin{cases}
\kappa_1/\Omega^d,\  \ \ \ \ \ \ \text{for } \Omega \to 0,\\
\kappa_2 (\pi-\Omega),\  \ \text{for } \Omega \to \pi,
\end{cases}
\end{eqnarray}
where $\kappa_1$ is the strip Casimir amplitude (\ref{Tij strip}),  and $\kappa_2$ is universally determined by the displacement operator \cite{Miao:2024ddp}
\begin{eqnarray}\label{sect6: universal relation} 
\kappa_2= \frac{d\pi^{\frac{d-2}{2}}\Gamma(\frac{d}{2})}{2(d-1) \Gamma[d+2]} C_D.
\end{eqnarray}
The bound of $\kappa_1$ (\ref{Casimir bound}) and the universal relation (\ref{sect6: universal relation}) suggest the following lower bound for wedge space
\begin{eqnarray}\label{sect6: wedge bound} 
(\frac{-f(\Omega)}{C_D})\ge \lim_{T\to -(d-1)} (\frac{-f(\Omega)}{C_D})_{\text{holo}}, \ \text{for} \ 0<\Omega\le \pi.
\end{eqnarray}
In the following, we derive the holographic lower bound above and verify it by free BCFTs. 

The gravity dual of wedge space is given by AdS soliton with the metric \cite{Miao:2024ddp}
\begin{eqnarray}\label{sect6: wedge metric}
ds^2=\frac{\frac{dz^2}{h(z)}+h(z) d\theta^2+\frac{dr^2+\sum_{\hat{i},\hat{j}=1}^{d-2} \eta_{\hat{i}\hat{j}} dy^{\hat{i}} dy^{\hat{j}}}{r^2}}{z^2},
\end{eqnarray}
where $h(z)=1-z^2-c_1 z^d$. It produces the expected Casimir effect (\ref{Tij wedge}) provided \cite{Miao:2024ddp}
\begin{eqnarray}\label{sect6: c1}
c_1=f(\Omega)=z_{\max }^{-d} \left(\text{sech}^2\left(\rho \right)-z_{\max }^2\right),
\end{eqnarray}
where $z_{\max }$ is the turning point. From NBC (\ref{sect2: NBC}) with $\lambda=0$, we can fix the opening angle of the wedge as \cite{Miao:2024ddp}
\begin{eqnarray}\label{sect6: opening angle result 1} 
\Omega_{\text{I}}=2\int_0^{z_{\text{max}}}\frac{dz}{h(z) \sqrt{\frac{h(z)}{h\left(z_{\max }\right)}-1}}, \ \text{for } T\le 0,
\end{eqnarray}
and 
\begin{eqnarray}\label{sect6: opening angle result 2} 
\Omega_{\text{II}}=\beta-\Omega_{\text{I}}(\rho \to -\rho), \ \text{for } T\ge 0.
\end{eqnarray}
Here $\beta$ is the angle period in bulk 
\begin{eqnarray}\label{dual: period}
\beta=\frac{4\pi}{|h'(z_h)|}=\frac{4\pi z_h}{d+(2-d) z_h^2},
\end{eqnarray}
with $h(z_h)=0$. From (\ref{sect6: c1},\ref{sect6: opening angle result 1},\ref{sect6: opening angle result 2}), we can derive $f(\Omega)$ for any given brane tension $T=(d-1)\tanh(\rho)$. $f(\Omega)$ is obtained exactly for $d=2, 4$ and numerically for general $d$ \cite{Miao:2024ddp}. The norm of the displacement operator reads \cite{Miao:2024ddp}
\begin{eqnarray}\label{sect6: CD}
C_D=\frac{2 (d-1) \Gamma (d+2)}{d \pi ^{\frac{d-2}{2}} \Gamma \left(\frac{d}{2}\right)}\begin{cases}
\frac{-d \Gamma \left(\frac{d}{2}\right) \text{sech}^3\left(\rho \right) \left(-\text{csch}\left(\rho \right)\right){}^{-d}}{\sqrt{\pi } \Gamma \left(\frac{d+1}{2}\right) \left(\text{sech}^3\left(\rho \right) F+d \text{csch}\left(\rho \right) \left(-\tanh \left(\rho \right)\right){}^d\right)}, \ \ \ \ \ \ \ \ \ \ \ \ \ \ \ \ \ \ \ \ \text{for } \rho\le 0,\\
\frac{d \Gamma \left(\frac{d}{2}\right)}{\pi  (d-1) d \Gamma \left(\frac{d}{2}\right)-\frac{\sqrt{\pi } \Gamma \left(\frac{d+1}{2}\right) \left(d \left(\tanh \left(\rho \right) \text{sech}\left(\rho \right)\right){}^d-\sinh \left(\rho \right) \text{sech}^{d+3}\left(\rho \right) F\right)}{\text{sech}^2\left(\rho \right) \tanh ^{d+1}\left(\rho \right)}},\ \ \text{for }  \rho\ge 0,
\end{cases}
\end{eqnarray} 
where $F=\, _2F_1\left(\frac{d-1}{2},\frac{d}{2};\frac{d+2}{2};-\text{csch}^2\left(\rho \right)\right)$. In the limit $x=\text{sech}^2(\rho)\to 0$, we have 
\begin{eqnarray}\label{sect6: CD limit}
C_D=\frac{2 (d-1) \pi ^{\frac{1}{2}-\frac{d}{2}} \Gamma (d+2)}{d \Gamma \left(\frac{d+1}{2}\right)} x^{1-\frac{d}{2}}+O\left(x^{2-\frac{d}{2}}\right).
\end{eqnarray}

\begin{figure}[t]
\centering
\includegraphics[width=10cm]{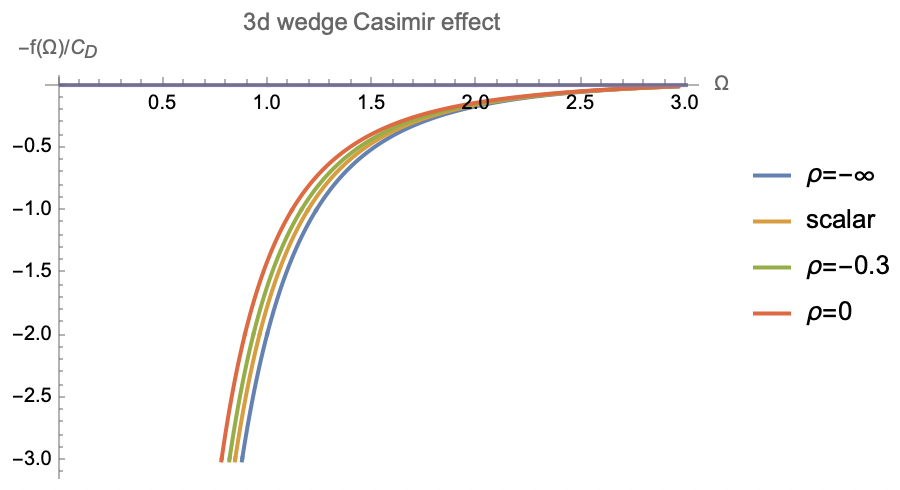} 
\caption{$(-f(\Omega)/C_D)$ for 3d BCFTs. The blue, green, red and yellow curves correspond to AdS/BCFT with $\rho=-\infty, -0.3,0$, and free scalar. It shows holography with $T\to -2$ ($\rho\to -\infty$) (blue curve) sets the lower bound of wedge Casimir effect.  }
\label{wedge bound 3d}
\end{figure}

\begin{figure}[t]
\centering
\includegraphics[width=10cm]{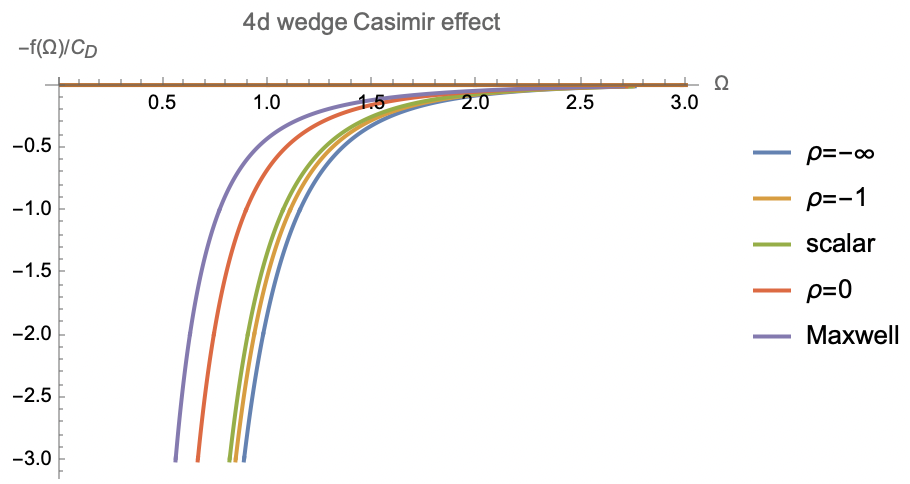} 
\caption{ $(-f(\Omega)/C_D)$ for 4d BCFTs. The blue, yellow, red, green, and purple curves correspond to AdS/BCFT with $\rho=-\infty, -1,0$, scalar and Maxwell field. It shows holography with $T\to -3$ ($\rho\to -\infty$) (blue curve) sets the lower bound of wedge Casimir effect.  }
\label{wedge bound 4d}
\end{figure}

Now let us consider the holographic lower bound (\ref{sect6: wedge bound}). Similar to the case of the strip, the lower bound  (\ref{sect6: wedge bound}) is achieved with the minimal brane tension $T\to -(d-1)$ ($\rho\to -\infty$).  To have a finite ratio $-f(\Omega)/C_D$, we consider the limit $x\to 0$, $f(\Omega) \to \infty$ with $\hat{f}(\Omega)=f(\Omega) x^{\frac{d}{2}-1}$ finite.  Then, the ratio becomes 
\begin{eqnarray}\label{sect6: ratio limit}
\text{ra}(\Omega)=\lim_{x\to 0}( \frac{-f(\Omega)}{C_D})=\lim_{x\to 0} \frac{-f(\Omega) x^{\frac{d}{2}-1}}{\frac{2 (d-1) \pi ^{\frac{1}{2}-\frac{d}{2}} \Gamma (d+2)}{d \Gamma \left(\frac{d+1}{2}\right)} }=\frac{-\hat{f}(\Omega)}{\frac{2 (d-1) \pi ^{\frac{1}{2}-\frac{d}{2}} \Gamma (d+2)}{d \Gamma \left(\frac{d+1}{2}\right)} }
\end{eqnarray} 
From (\ref{sect6: c1}), we solve
\begin{eqnarray}\label{sect6: zmax}
z_{\max}=z_0 \sqrt{x}, \ \text{with} \ \hat{f} z_0^d+z_0^2-1=0.
\end{eqnarray} 
Substituting (\ref{sect6: zmax}) together with $c_1=f=\hat{f} x^{1-\frac{d}{2}}$ into (\ref{sect6: opening angle result 1}) and taking the limit $x\to 0$, we obtain 
\begin{eqnarray}\label{sect6: opening angle result limit} 
\Omega=\int_0^{1}dy \frac{2 z_0}{\sqrt{1-y^2 z_0^2+\left(z_0^2-1\right) y^d}},
\end{eqnarray}
where $z_0$ is a function of $\hat{f}$ from (\ref{sect6: zmax}). We remark that (\ref{sect6: opening angle result limit}) gives the correct limit (\ref{sect6: limits}) for $f=\hat{f}=0$ ($z_0=1$)
\begin{eqnarray}\label{sect6: opening angle result limit 1} 
\lim_{f\to 0}\Omega=\int_0^{1}dy \frac{2}{\sqrt{1-y^2 }}=\pi,
\end{eqnarray}
which is a test of our calculations. For $d=2, 4$, we derive exact expressions 
\begin{eqnarray}\label{sect6: angle 2d 4d}
\Omega=\begin{cases}
\pi  z_0, \ \ \ \ \ \  \ \ \ \ \ \ \ \ \ \ \text{for } d=2,\\
2 z_0 K\left(z_0^2-1\right),\ \ \text{for } d=4,
\end{cases}
\end{eqnarray} 
where $K$ denotes the complete elliptic integral of the first kind. From (\ref{sect6: ratio limit}) and (\ref{sect6: zmax}), we can express $z_0$ in terms of the ratio $\text{ra}$ (\ref{sect6: ratio limit}) and then rewrite (\ref{sect6: angle 2d 4d}) as
\begin{eqnarray}\label{sect6: ratio 2d 4d}
\Omega=\begin{cases}
\frac{\pi ^{3/2}}{\sqrt{\pi -12 \text{ra}}}, \ \ \ \ \ \  \ \ \ \ \ \ \ \ \ \  \ \ \ \ \ \  \ \ \ \ \ \ \ \ \ \  \ \ \ \ \ \  \ \ \ \ \ \ \ \ \ \ \ \ \ \text{for } d=2,\\
\frac{1}{2} \sqrt{\frac{\pi }{30}} \sqrt{\frac{\pi -\sqrt{\pi ^2-960 \text{ra}}}{\text{ra}}} K\left(\frac{\pi  \left(\pi -\sqrt{\pi ^2-960 \text{ra}}\right)}{480 \text{ra}}-1\right),\ \ \text{for } d=4.
\end{cases}
\end{eqnarray} 
Now we have obtained the exact relations between the ratio $\text{ra}=(-f(\Omega)/C_D)$ and the opening angle $\Omega$ in the limit $T\to -(d-1)$ for $d=2,4$. For general dimensions, we can derive $\text{ra}(\Omega)$ numerically from (\ref{sect6: ratio limit},\ref{sect6: zmax},\ref{sect6: opening angle result limit}). To end this section, we draw $(-f(\Omega)/C_D)$ for free and holographic BCFTs in Fig. \ref{wedge bound 3d} for $d=3$ and Fig. \ref{wedge bound 4d} for $d=4$. The expressions of $f(\Omega)$ for free BCFTs can be found in \cite{Miao:2024ddp}. Fig. \ref{wedge bound 3d} and Fig. \ref{wedge bound 4d} show the AdS/BCFT with the minimal tension $T\to -(d-1)$ imposes the lower bound of wedge Casimir effect. The results of this section imply that holography sets a lower bound of the Casimir effect for general boundary shapes, not just for the strip. We leave the study of general boundary shapes to future work. We remark that the holographic lower bound for the wedge Casimir effect is independent of the gravity models. We have checked that DGP gravity in the normal phase yields the same results as Einstein gravity. 

\section{Conclusions and Discussions}

In this paper, we explore the fundamental bound of the Casimir effect in general dimensions. We propose holography impose a universal lower bound on the Casimir coefficient ratio to the displacement operator norm. It is similar to the famous KSS bound \cite{Kovtun:2004de} for hydromechanics and sheds light on the crucial question: how large the Casimir effect one could produce in principle. We derive the universal bound for a strip by studying various holographic models and verify it with free BCFTs and the $O(N)$ model in $\epsilon$ expansions. We also derive the holographic bound of the Casimir effect for a wedge and test it with free BCFTs. Our proposal is expected to work for general boundary shapes, not limited to the strip and wedge.  

There are many essential problems worth exploring. 

\begin{itemize}
  \item It is interesting to find a field-theoretical proof or counterexample of the holographic bounds (\ref{Casimir bound},\ref{wedge bound}). The conformal bootstrap is a powerful tool for achieving this target. 

  \item This paper focuses on the Casimir effect of the strip and wedge. It is interesting to generalize the discussions to general boundary shapes. Near a curved boundary, the Casimir effect takes a universal form determined by Weyl anomaly \cite{Miao:2017aba}
\begin{eqnarray}\label{sect4: Tij from Weyl anomaly}
\langle T_{ij} \rangle  = \alpha_1 \frac{\bar{k}_{ij}}{x^{d-1}}+..., \ \ x\sim 0,
\end{eqnarray}
where $\bar{k}_{ij}$ are the traceless parts of extrinsic curvatures, $x$ is the distance to the boundary, and $\alpha_1$ is boundary central charge of Weyl anomaly, or equivalently, the norm of displacement operator \cite{Miao:2018dvm}
 \begin{eqnarray}\label{sect4: universalrelation1}
\alpha_1=-\frac{d\Gamma[\frac{d+1}{2}]\pi^{\frac{d-1}{2}}}{(d-1)\Gamma[d+2]}C_D.
\end{eqnarray}
We get finite terms by subtracting the divergent terms (\ref{sect4: Tij from Weyl anomaly}) from the Casimir effect. The ratios of these finite terms and $C_D$ are expected to have a universal lower bound. We leave this non-trivial problem to future work. 

  \item For simplicity, this paper focuses on BCFTs. It is interesting to generalize our proposal to non-BCFTs. Let us consider the massive free theories. For non-BCFTs, we define the norm of displacement operator in the short-distance limit
 \begin{eqnarray}\label{CD for nonBCFTs} 
\langle D(y) D(0) \rangle =\frac{C_D}{|y|^{2d}}+O(|y|^{2d-1}).
\end{eqnarray}
By dimensional analysis, $C_D$ is independent of the mass. It agrees that the mass effect can be ignored at UV, i.e., $m y\ll 1$. On the other hand, the Casimir effect is suppressed by the mass \cite{Bordag:2009zz}. As a result, the ratios, such as $(-\kappa_1/C_D)$ and $(-f(\Omega)/C_D)$, increase with the mass and thus obey the lower bounds (\ref{Casimir bound},\ref{wedge bound}) of Casimir effect. Exploring the scope within which the holographic bound is obeyed for non-BCFT is interesting. 
  
  \item For simplicity, we impose the same boundary conditions on the two boundaries of the strip and wedge. It is interesting to generalize to the case of mixed boundary conditions, which can be realized by adding matter fields on the branes. 
    
 \end{itemize}

\section*{Acknowledgements}

We thank T.~Takayanagi, J. X. Lu, S. Lin, W. L. Li, Z. J. Li, N. Su for valuable comments and discussions. We are grateful to W. z. Guo, L. Y. Huang, S. He, S. M. Ruan and many participants of ``Gauge Gravity Duality 2024" for helpful discussions. This work is supported by the National Natural Science Foundation of China (No.12275366).

\appendix

\section{Ghost-free condition for GB-DGP gravity}

Following the approach of \cite{Miao:2023mui}, we derive the ghost-free condition (\ref{sect3: ghost-free condition}) for GB-DGP gravity (\ref{sect3: IGBmiao}) with $d=4$ in this appendix. We take the following ansatz of the perturbation metric and the embedding function of the brane
 \begin{eqnarray}\label{appA: perturbationmetric}
&&ds^2=dr^2+\cosh^2 (r) \left( \bar{h}^{(0)}_{ij}(y) + \epsilon H(r) \bar{h}^{(1)}_{ij}(y)  \right)dy^i dy^j+O(\epsilon^2),\\
&& Q: \ r=\rho+O(\epsilon^2), \label{perturbationQ}
\end{eqnarray}
where $\bar{h}^{(0)}_{ij}$ is the AdS metric with a unit radius and $\bar{h}^{(1)}_{ij}$ denotes the perturbation.  Under the transverse traceless gauge 
 \begin{eqnarray}\label{appA: hij1gauge}
\bar{D}^i \bar{h}^{(1)}_{ij}=0,\ \ \  \bar{h}^{(0)ij}\bar{h}^{(1)}_{ij}=0,
\end{eqnarray}
the equations of motion (EOM) of GB-DGP gravity can be separated variables as
\begin{eqnarray}\label{appA: EOMMBCmassivehij}
&& \left(\bar{\Box}+2-m^2\right)\bar{h}^{(1)}_{ij}(y)=0,\\
&& \cosh^2(r) H''(r)+4 \sinh (r) \cosh (r) H'(r) + m^2 H(r)=0, \label{appA: EOMMBCmassiveH}
\end{eqnarray}
where $m$ is the graviton mass. We impose DBC on the AdS boundary while NBC (\ref{sect3: NBC}) on the brane, which yields
\begin{eqnarray}\label{appA: DBC for H}
&& H(-\infty)=0,\\
&&H'(\rho )= 2 m^2 H(\rho ) \text{sech}^2(\rho )\Big(\frac{\lambda+12 \alpha  \lambda +2 \alpha  \tanh (\rho ) }{1+8 \alpha} \Big). \label{appA: NBC for H}
\end{eqnarray}
From EOM (\ref{appA: EOMMBCmassiveH}) and BCs (\ref{appA: DBC for H},\ref{appA: NBC for H}), we derive the orthogonal relationship for KK modes
 \begin{eqnarray}\label{appA: orthogonal-gravity}
\langle H_m, H_{m'} \rangle&=&c_m\delta_{m, m'}=\int_{-\infty}^{\rho}\frac{\cosh^{2}(r)}{\cosh ^{2}(\rho )}H_m(r) H_{m'}(r) dr\nonumber\\
&+&2 \Big(\frac{\lambda+12 \alpha  \lambda +2 \alpha  \tanh (\rho ) }{1+8 \alpha} \Big)H_m(\rho) H_{m'}(\rho).
\end{eqnarray}
To have a positive inner product $c_m=\langle H_m, H_{m} \rangle$, both the bulk integral and boundary term of (\ref{appA: orthogonal-gravity}) should be positive, which gives the ghost-free condition (\ref{sect3: ghost-free condition}). Let us provide solid proof of this statement below. 

Defining the step function 
\begin{eqnarray}\label{appA: step function 0}
\Pi_0(r)=\begin{cases}
0,\  \ \ \ \ \ \ \text{for } r<\rho,\\
1,\  \ \ \ \ \ \ \text{for } r\ge \rho.
\end{cases}
\end{eqnarray}
and expending it in terms of KK modes, we obtain
\begin{eqnarray}\label{appA: Pi0 expand}
\Pi_0(r)=\sum_m \frac{\langle \Pi_0, H_m \rangle}{\langle H_m, H_m \rangle } H_m(r)=2\Big(\frac{\lambda+12 \alpha  \lambda +2 \alpha  \tanh (\rho ) }{1+8 \alpha} \Big) \sum_m  \frac{H_m(\rho) H_m(r)}{\langle H_m, H_m \rangle},
\end{eqnarray}
where we have used $\langle \Pi_0, H_m \rangle=2\Big(\frac{\lambda+12 \alpha  \lambda +2 \alpha  \tanh (\rho ) }{1+8 \alpha} \Big) H_m(\rho)$ derived from (\ref{appA: orthogonal-gravity},\ref{appA: Pi0 expand}). Noting that $\Pi_0(\rho)=1$, the above equation yields the so-called spectrum identity
\begin{eqnarray}\label{appA: spectrum identity}
\sum_m  \frac{H_m(\rho) H_m(\rho)}{\langle H_m, H_m \rangle}=\frac{1}{2\Big(\frac{\lambda+12 \alpha  \lambda +2 \alpha  \tanh (\rho ) }{1+8 \alpha} \Big) }.
\end{eqnarray}
For a ghost-free theory with $\langle H_m, H_m \rangle>0$, the left-hand side of (\ref{appA: spectrum identity}) is positive. As a result, the right-hand side of (\ref{appA: spectrum identity}) must also be positive, which gives the ghost-free condition (\ref{sect3: ghost-free condition}) 
\begin{eqnarray}\label{appA: ghost-free condition}
\frac{\lambda+12 \alpha  \lambda +2 \alpha  \tanh (\rho ) }{1+8 \alpha} \ge 0.
\end{eqnarray}

According to \cite{Miao:2023mui}, the KK modes are automatically tachyon-free under the ghost-free condition (\ref{appA: ghost-free condition}). First, let us prove the mass spectrum is real. If there were complex $m^2$, they must appear in a complex conjugate pair, since both the EOM (\ref{appA: NBC for H}) and BCs (\ref{appA: DBC for H},\ref{appA: NBC for H}) are real. Then the orthogonal condition (\ref{appA: orthogonal-gravity}) becomes 
 \begin{eqnarray}\label{appA: orthogonal-gravity 1}
\langle H_m, H_{m^*} \rangle=\int_{-\infty}^{\rho}\frac{\cosh^{2}(r)}{\cosh ^{2}(\rho )} |H_m(r)|^2 dr+2 \Big(\frac{\lambda+12 \alpha  \lambda +2 \alpha  \tanh (\rho ) }{1+8 \alpha} \Big) |H_m(\rho)|^2 > 0.
\end{eqnarray}
On the other hand, we have $\langle H_m, H_{m^*} \rangle=c_m \delta_{m,m^*}=0$. The contradiction means there are no complex masses for the KK modes. Let us go on to prove $m^2>0$. From EOM (\ref{appA: NBC for H}) and BCs (\ref{appA: DBC for H},\ref{appA: NBC for H}), we derive
\begin{eqnarray}\label{appA: positive mass 1}
&&\int_{-\infty}^{\rho} \cosh^{4} (r)H'_m(r)H'_m(r)  dr=\cosh^{4} (\rho)  H'_m(\rho) H_m(\rho)-\int_{-\infty}^{\rho} (\cosh^{4} (r)H'_m(r))'  H_{m}(r)  dr\nonumber\\
&=&m^2 \Big[\int_{-\infty}^{\rho}\cosh^{2}(r)H^2_m(r) dr+2\Big(\frac{\lambda+12 \alpha  \lambda +2 \alpha  \tanh (\rho ) }{1+8 \alpha} \Big) \cosh^{2}(\rho)  H^2_m(\rho) \Big]\nonumber\\
&=& m^2 \langle H_m, H_{m} \rangle \cosh^{d-2}(\rho).
\end{eqnarray}
Because $\int_{-\infty}^{\rho} \cosh^{4} (r)H'_m(r)H'_m(r)  dr$ and $\langle H_m, H_{m} \rangle$ are both positive, the above equation leads to $m^2>0$.  

We have obtained the ghost-free condition (\ref{appA: ghost-free condition}) and prove it automatically yields a tachyon-free mass spectrum. Note that the brane tension can be negative in the above discussions. For example, $\rho$ can be negative in the orthogonal condition (\ref{appA: orthogonal-gravity}) and all the conclusions derived from it. A quick way to see this is by considering Einstein's gravity with $\lambda=\alpha=0$. Thus, we confirm that the negative brane tension is well-defined in AdS/BCFT, as we stressed in the main text. 

\section{Displacement operator for GB-DGP gravity}

By applying the method of \cite{Miao:2017aba}, we derive the norm of displacement operator for GB-DGP gravity (\ref{sect3: ghost-free condition}) with $d=4$ in this appendix. We consider the perturbative metric of a half-space
\begin{eqnarray}\label{appB: bulkmetric}
&& ds^2=\frac{1}{z^2}\Big{[} dz^2+dx^2 
+\Big(\delta_{ab}-2 \epsilon x \bar{k}_{ab} f(\frac{z}{x})
 \Big)dy^a dy^b+O(\epsilon^2)\Big{]},
\end{eqnarray}
and the embedding function of brane $Q$
\begin{eqnarray}\label{appB: Q}
x=-\sinh\rho \ z + O(\epsilon^2),
\end{eqnarray}
where $\epsilon$ denotes the expanding order. 
We impose DBC $f(0)=1$ on the AdS boundary $z=0$ so that $ \bar{k}_{ab}$ become the traceless parts of extrinsic curvatures for BCFTs. Substituting (\ref{appB: bulkmetric}) into EOM of GB-DGP gravity, we get one independent equation 
\begin{eqnarray}\label{appB: EOM}
s \left(s^2+1\right) f''(s)-3 f'(s)=0, 
\end{eqnarray}
which can be solved as 
\begin{eqnarray}\label{appB: solution}
f(s)=1+2 d_1-d_1\frac{ \left(2+s^2\right)}{\sqrt{1+s^2}}.
\end{eqnarray}
Above, we have used $f(0)=1$ to fix one integral constant. Substituting (\ref{appB: solution}) and (\ref{appB: bulkmetric},\ref{appB: Q}) into NBC (\ref{sect3: NBC}), we determine the other integral constant
\begin{eqnarray}\label{appB: integral constant}
d_1=-\frac{e^{\rho } \cosh (\rho ) (4 (12 \alpha +1) \lambda  \sinh (2 \rho )+(16 \alpha +1) \cosh (2 \rho )+1)}{e^{4 \rho } (16 \alpha  (3 \lambda +1)+4 \lambda +1)+e^{2 \rho } (8 \alpha  (6 \lambda +1)+4 \lambda +2)+8 \alpha +1}.
\end{eqnarray}
From (\ref{appB: bulkmetric},\ref{appB: solution}), we derive the holographic stress tensor for GB gravity \cite{Sen:2014nfa}
\begin{eqnarray}\label{appB: holo Tij}
\langle T_{ab} \rangle  = 2 (1+8 \alpha)d_1  \frac{\epsilon \bar{k}_{ab}}{x^3}+O(\epsilon^2).
\end{eqnarray}
Comparing (\ref{appB: holo Tij}) with (\ref{sect4: Tij from Weyl anomaly}), we read off $2 (1+8 \alpha)d_1=\alpha_1$. Then from (\ref{sect4: universalrelation1}) and (\ref{appB: integral constant}), we finally obtain the norm of displacement operator (\ref{sect3: CD 4d}) for GB-DGP gravity
\begin{eqnarray}\label{appB: displacement operator}
C_D=\frac{240 (8 \alpha +1) e^{\rho } \cosh (\rho ) (4 (12 \alpha +1) \lambda  \sinh (2 \rho )+(16 \alpha +1) \cosh (2 \rho )+1)}{\pi ^2 \left(e^{4 \rho } (16 \alpha  (3 \lambda +1)+4 \lambda +1)+e^{2 \rho } (8 \alpha  (6 \lambda +1)+4 \lambda +2)+8 \alpha +1\right)}.
\end{eqnarray}


\end{document}